\documentclass[twocolumn,preprintnumbers,amsmath,amssymb]{revtex4}

\usepackage{graphicx}
\usepackage{dcolumn}
\usepackage{bm}
\usepackage{rotating}
\usepackage{verbatim}

\begin{document}

\title{Comparison of secondary islands in collisional reconnection to
Hall reconnection}



\author{L.~S.~Shepherd and P.~A.~Cassak} 
\affiliation{Department of Physics,
West Virginia University, Morgantown, WV, 26506, USA}


\begin{abstract}
Large-scale resistive Hall-magnetohydrodynamic (Hall-MHD)
simulations of the transition from Sweet-Parker (collisional) to Hall
(collisionless) magnetic reconnection are presented, the first to
separate effects of secondary islands from collisionless effects.
Three main results are described.  There exists a regime in which
secondary islands occur without collisionless effects when the
thickness of the dissipation regions exceed ion gyroscales.  The
reconnection rate with secondary islands is faster than Sweet-Parker
but significantly slower than Hall reconnection.  This implies that
secondary islands are not the cause of the fastest reconnection rates.
Because Hall reconnection is much faster, its onset causes the
ejection of secondary islands from the vicinity of the X-line.  These
results imply that most of the energy release occurs during Hall
reconnection.  Coronal applications are discussed.
\end{abstract}

\preprint{Submitted to {\it Phys.~Rev.~Lett.}, May 11, 2010}

\maketitle


Magnetic reconnection is widely regarded as the mechanism underlying
energy release in the solar corona during flares, coronal mass
ejections, and coronal jets \cite{Birn07}.  The Sweet-Parker model
\cite{Sweet58,Parker57} was the first self-consistent theory, but is
far too slow to explain observations.  Much has been invested in
faster reconnection scenarios, such as collisionless (Hall)
reconnection \cite{Birn01} in which the Hall term plays a key role
\cite{Malakit09,Cassak10}.  Hall reconnection seems fast enough to
explain observed energy release rates \cite{Shay99}.  Lately, the role
of secondary islands (plasmoids) on Sweet-Parker reconnection has
generated much interest.  While they were discussed some time ago
\cite{Matthaeus85,Matthaeus86,Biskamp86}, systematic studies were not
carried out until recently.  It has been argued in various contexts
that secondary islands make reconnection faster than Sweet-Parker
reconnection \cite{Matthaeus85,Kliem95,Lazarian99,Lapenta08}.  (Note,
we are discussing secondary islands occurring during collisional
reconnection, not those that occur after collisionless reconnection
has begun \cite{Daughton06}.)

Understanding secondary islands in Sweet-Parker reconnection is
important for explaining coronal evolution.  On the theoretical side,
the reconnection rate places constraints on the dynamics.  For
example, if secondary islands make Sweet-Parker reconnection much
faster or hasten the transition to fast reconnection, it cannot take
place during pre-flare energy storage.  If it remains slow, then it
can occur while energy accumulates \cite{Cassak05,Uzdensky07,
Cassak08a}.  On the observational side, it was hypothesized that high
density blobs in current sheets during solar eruptions are secondary
islands \cite{Ciaravella08,Lin09}.  Also, numerous observations of
reconnection processes display a slow phase preceding an eruptive
event with an abrupt transition.  Examples include non-eruptive flux
emergence \cite{Longcope05} and flows during coronal implosion as a
result of an impulsive flare \cite{Liu09,Liu10}.

Past work on secondary islands showed that they appear spontaneously
due to a secondary tearing instability when the Lundquist number $S =
4 \pi c_{A} L_{SP} / \eta c^{2}$ exceeds $\sim 10^{4}$
\cite{Biskamp86}, where $L_{SP}$ is the half-length of the
Sweet-Parker dissipation region, $\eta$ is the resistivity, and
$c_{A}$ is the Alfv\'en speed based on the reconnecting magnetic
field.  Equivalently, this can be written as $\delta / L_{SP} < 0.01$,
where $\delta$ is the thickness of the dissipation region.  A study of
the linear phase of the instability \cite{Loureiro07} found a growth
rate faster than the Alfv\'en transit time along the sheet.  Recent
simulations addressed the nonlinear reconnection rate $E$ for high
$S$, showing that it is considerably faster than the Sweet-Parker rate
and its dependence on $S$ becomes weak
\cite{Bhattacharjee09,Cassak09a}.  However, the simulations only go up
to $S \sim 10^{6}$, so $E$ is only one order of magnitude faster than
the Sweet-Parker rate and it is not clear whether it will be fast or
slow at larger $S$.  Other relevant studies showed that $E$ increases
with the square root of the number of islands
\cite{Daughton09,Cassak09a} and that secondary islands are suppressed
when reconnection is embedded, meaning that the upstream field is
smaller than the asymptotic field \cite{Cassak09b}.  Many studies
consider secondary islands caused by external random magnetic
perturbations \cite{Smith04,Kowal09,Loureiro09,Skender10}.  Other
studies include the interaction of multiple islands \cite{Nakamura10}
and a statistical model of multiple island interaction \cite{Fermo10}.

In addition to increasing the reconnection rate, secondary islands
hasten the transition to Hall reconnection \cite{Shibata01,
Daughton09}.  When a secondary island forms, the fragmented current
sheet is shorter, so its Sweet-Parker thickness is smaller
\cite{Daughton09,Cassak09a}.  When the layer reaches ion gyroscales
\cite{Mandt94,Ma96}, Hall reconnection begins abruptly
\cite{Bhattacharjee04,Cassak05,Cassak06,Cassak07c,Daughton09}.  This
was recently verified using collisional particle-in-cell (PIC)
simulations \cite{Daughton09,Daughton09b}.

The only previous study to include both secondary islands and the Hall
effect utilized large PIC simulations \cite{Daughton09,Daughton09b},
but numerical constraints forced $S$ to be small enough that Hall
reconnection began as soon as a secondary island formed.  Since the
Hall effect arises only at ion gyroscales, there should be a regime in
which secondary islands are present without the Hall effect playing a
role if the sheet is thicker than ion gyroscales.  The goal of this
study is to separate the two effects and ascertain which one leads to
dramatically larger reconnection rates, which dictates the mechanism
releasing the majority of the energy during the eruptive phase of
reconnection.

In this paper, the first simulations to separate the two effects are
presented.  There are three main results: (1) there is a regime in
which secondary islands occur without collisionless effects entering,
(2) the reconnection rate is faster than Sweet-Parker but
significantly slower than Hall reconnection, which shows that
secondary islands are not the cause of the highest reconnection rates,
and (3) the onset of Hall reconnection ejects secondary islands in the
vicinity of the X-line.  The latter two imply that the majority of the
energy release occurs at Hall reconnection sites.  Coronal
applications are discussed.


Numerical simulations are performed using the two-fluid code F3D
\cite{Shay04}.  Magnetic fields and densities are normalized to
arbitrary values $B_{0}$ and $n_{0}$, velocities to the Alfv\'en speed
$c_{A0} = B_{0} / (4 \pi m_{i} n_{0})^{1/2}$ where $m_{i}$ is the ion
mass, lengths to the ion inertial length $d_{i0} = c / \omega_{pi}$,
times to the ion cyclotron time $\Omega_{ci}^{-1}$, electric fields to
$E_{0} = c_{A0} B_{0} /c$, and resistivities to $\eta_{0} = 4 \pi
c_{A0} d_{i0} / c^{2}$.

The initial configuration is a double tearing mode with two Harris
sheets, $B_{x}(y) = \tanh[(y + L_{y}/4) / w_{0}] - \tanh[(y - L_{y}/4)
/ w_{0}] - 1$, where $w_{0}$ is the initial current layer thickness
and $L_{y}$ is the system size in the inflow direction.  Total
pressure is balanced initially using a non-uniform density which
asymptotes to 1.  The temperature $T = 1$ is constant and uniform.  A
single X-line is seeded using a coherent magnetic perturbation of
amplitude $1.6 \times 10^{-2}$ to rapidly achieve nonlinear
reconnection.  Initial random magnetic perturbations break symmetry so
secondary islands are ejected.  There is no initial out-of-plane
(guide) magnetic field.  Boundaries in both directions are periodic.
Electron inertia is $m_{e} = m_{i} / 25$.  This value is acceptable
since we focus on the onset of Hall reconnection at ion scales rather
than electron scales.

Simulation parameters are chosen so reconnection will proceed in three
distinct phases: Sweet-Parker without secondary islands, Sweet-Parker
with secondary islands, and Hall reconnection.  A very large system
size $L_{x} \times L_{y} = 819.2 \times 409.6$ is employed with
resistivity $\eta = 0.008$, corresponding to a global Lundquist number
$S_{g} = L_{x} / \eta \sim 10^{5}$, which exceeds the Biskamp
criterion of $10^{4}$.  To postpone secondary island onset, we choose
$w_{0} = 12.0$ which makes the reconnection embedded \cite{Cassak09b}.
Embedding makes the Sweet-Parker layer thicker since $\delta \sim
(\eta L_{SP} / c_{A up})^{1/2}$, where $c_{A up}$ is the Alfv\'en
speed based on the upstream magnetic field $B_{up}$.  For wide current
layers, $B_{up} \sim B_{0} \delta / w_{0}$ \cite{Cassak09b}, so
eliminating $B_{up}$ gives $\delta \sim (\eta L_{SP} w_{0})^{1/3} \sim
2.7$, where $L_{SP} \sim L_{x} / 4 \sim 200$ in our periodic system.
Thus, the layer begins wider than $d_{i}$, and since $\delta / L_{SP}
> 0.01$, no secondary islands occur initially and the system undergoes
Sweet-Parker reconnection.  The reconnection inflow convects in
stronger magnetic fields, so the current sheet self-consistently
thins.  Islands arise when $\delta / L_{SP} \sim 0.01$, which gives
$\delta \sim 2.0$ since $L_{SP} \sim 200$.  It has been shown
\cite{Daughton09,Cassak09a} that if $N$ X-lines are present, $\delta$
decreases by a factor of $N^{1/2}$.  For a single secondary island ($N
= 2$), the layer shrinks to $\delta \sim 2.0 / 2^{1/2} \sim 1.4$.
This exceeds $d_{i}$, so Sweet-Parker with secondary islands should
persist.  Hall reconnection only starts when $\delta \sim 1$, so three
distinct phases occur.

A simulation is first performed with a grid scale $\Delta = 0.2$ and
the results are qualitatively consistent with expectations.  To assure
$\Delta$ does not play a role, the simulations are redone with $\Delta
= 0.1$, giving comparable results.  Data is presented only from the
high resolution runs.  The equations employ fourth order diffusion
with coefficient $D_{4} = 1.75 \times 10^{-4}$ to damp noise at the
grid scale.  A smaller value of $D_{4}$ leads to a slightly larger
Hall reconnection rate, but does not alter our key conclusions.

\begin{figure}
\includegraphics[width=3.4in]{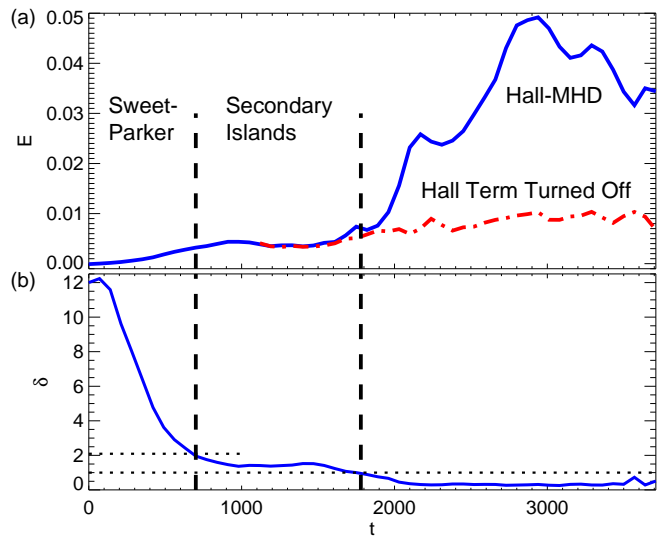}
\caption{\label{fig-data} (Color) (a) Reconnection rate $E$ as a
function of time $t$.  The solid (blue) line is a Hall-MHD run.
Dashed lines at $t \sim 700$ and 1780 indicate the onset of secondary
islands and Hall reconnection, respectively.  The dot-dashed (red)
line shows $E$ for a simulation restarted at $t = 1120$ with no Hall
effect and $m_{e} = 0$.  (b) Thickness $\delta$ of the dissipation
region vs.~$t$.  Horizontal dotted lines mark predicted $\delta$ for
the onset of secondary islands ($\delta \sim 2$) and Hall reconnection
($\delta \sim 1$).}
\end{figure}

We now summarize the simulation results, followed by a careful
justification of the conclusions.  At early times, Sweet-Parker
reconnection prevails.  A secondary island first appears at $t \simeq
700$.  Reconnection proceeds with the secondary island until $t \simeq
1780$, when Hall reconnection onsets.  Thus, reconnection proceeds in
three distinct phases including an extended phase with secondary
islands but without the Hall effect triggered.

We compare the reconnection rate $E$ in the three phases to each other
and to theoretical predictions in Fig.~\ref{fig-data}(a), showing $E$
vs.~time $t$ as the solid (blue) line.  We measure $E$ as the time
rate of change in the difference in magnetic flux function $\psi$
between the main X-line and O-line.  Dashed lines at $t = 700$ and
1780 denote where secondary islands and the Hall effect arise,
respectively.  Ignoring secondary islands, Sweet-Parker theory
predicts $E \sim E_{SP} \sim 0.006$, where $E_{SP} \sim (\eta /
L_{SP})^{1/2}$, and $L_{SP} \sim 200$.  This assumes the magnetic
field is its asymptotic value of 1.  The measured value is $E \sim
0.004$, slightly lower than predicted as expected because $B_{up} < 1$
(the reconnection is embedded).  When $N$ X-lines are present, $E$
scales as $E_{SI} \sim E_{SP} \sqrt{N}$ \cite{Daughton09,Cassak09b}.
The measured rate of $0.005$ is consistent with this for a single
secondary island ($N = 2$).  After the Hall effect onsets, $E$
increases by about an order of magnitude.  Therefore, the reconnection
rate with secondary islands is faster than Sweet-Parker, but
significantly slower than Hall reconnection.

The transitions occur when predicted, as shown in
Fig.~\ref{fig-data}(b).  We plot $\delta$, measured as the half-width
at half-max of $J_{z}$ in the inflow direction through the X-line,
vs.~$t$.  The dotted lines at $\delta \sim 2$ and $1$ show the
predicted value when islands and the Hall effect should appear,
respectively.  These conditions are met at $t \simeq 700$ and 1780, in
good agreement with the observed transitions.

\begin{figure}
\includegraphics[width=3.4in]{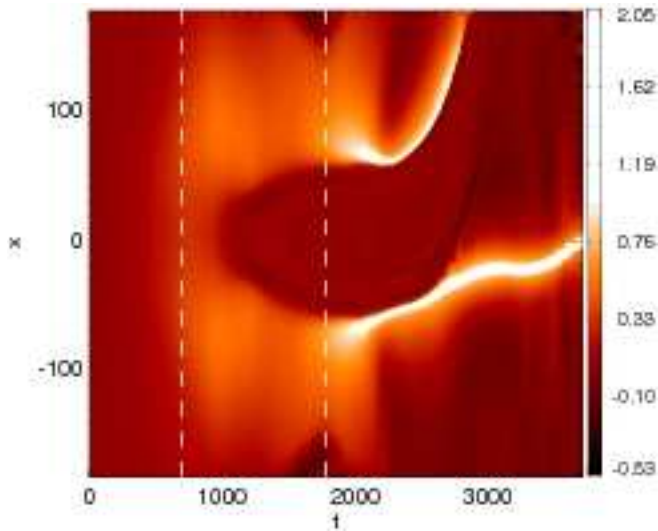}
\caption{\label{fig-data1} (Color) Time history plot of the
out-of-plane current density $J_{z}$ in the outflow direction.  Dashed
lines mark when a secondary island appears and when the Hall term
onsets.}
\end{figure}

The appearance of new physics can be seen in direct observations of
the out-of-plane current density $J_{z}$.  A two-dimensional time
history plot of $J_{z}$ in the outflow direction is plotted in
Fig.~\ref{fig-data1}.  Only the half domain centered on the seeded
X-line is shown.  The raw data is sampled at a rate of one frame per
70 time units, so linear interpolation is used to smooth data between
time slices.  The effect is cosmetic, not substantive.  The color bar
is stretched to enhance visibility of weaker currents.

Early in time, $J_{z}$ is structureless and extends the half-length of
the domain, as expected during Sweet-Parker reconnection.  A secondary
island near $x = 0$ appears as a dark spot with associated
strengthening of the fragmented current sheets.  This occurs at $t
\sim 700$, marked by the vertical dashed line.  This agrees with
Biskamp's criterion shown in Fig.~\ref{fig-data}(b).  As time evolves,
the island grows and $\delta$ shrinks.  When $\delta \sim d_{i}$, Hall
reconnection onsets and the current sheet becomes much shorter and
intense, appearing as a sharp peak in $J_{z}$ in Fig.~\ref{fig-data1}.
This begins at $t \sim 1780$, as also marked in
Fig.~\ref{fig-data}(b).

\begin{figure}
\includegraphics[width=3.4in]{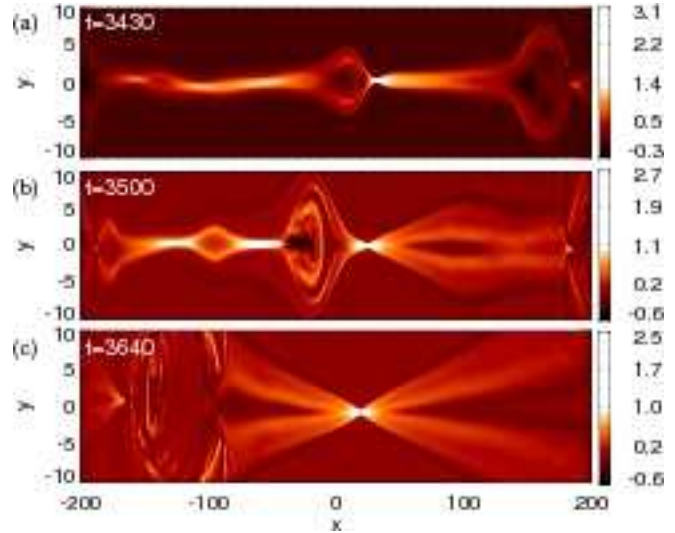}
\caption{\label{fig-data3} (Color) Time evolution of $J_{z}$ from the
other current sheet in our double tearing mode setup, showing the
ejection of secondary islands when Hall reconnection onsets. }
\end{figure}

There are two locations where Hall reconnection onsets.  An X-line
near $x \simeq -70$ onsets slightly earlier than an X-line at $x
\simeq 70$.  As Fig.~\ref{fig-data1} vividly shows, the latter X-line
is ejected from the dissipation region, along with the secondary
island, which is ejected at the Alfv\'en speed.  The ejection of the
secondary island implies that the two effect will not (locally)
coexist, so most of the energy is released at Hall reconnection sites.

This current sheet has only a single secondary island and one may ask
whether this result remains valid in more realistic settings with
multiple islands.  To address this, we show results from the other
current sheet in our double tearing mode setup, which
self-consistently develops multiple islands.  Figure \ref{fig-data3}
shows $J_{z}$ at three times near the onset of Hall reconnection.
Panel (a) is just as Hall reconnection onsets at $x \simeq 20$,
showing three pre-existing secondary islands.  The Hall reconnection
X-line grows steadily, as shown in panel (b).  Panel (c) shows that
the single X-line at $x \simeq 20$ is the only one to persist as all
of the secondary islands are ejected.  This suggests that the ejection
of nearby secondary islands by Hall reconnection sites is a robust
result, and may reasonably represent local behavior in a macroscopic
current sheet.

\begin{figure}
\includegraphics[width=3.4in]{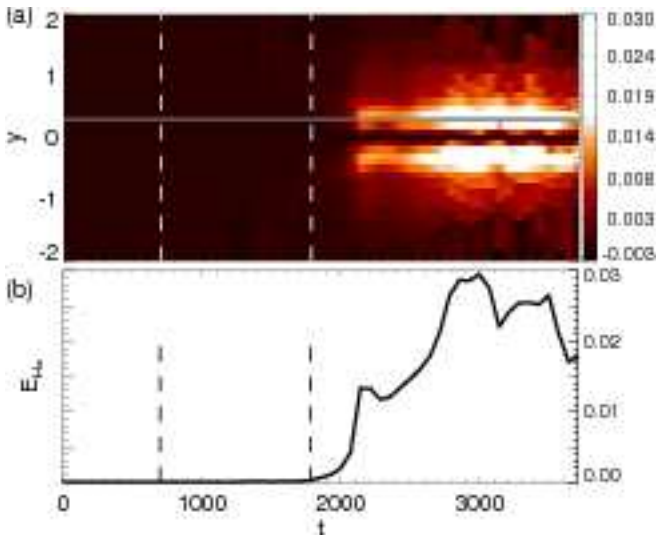}
\caption{\label{fig-data2} (Color) (a) Time history plot of the
out-of-plane Hall electric field $E_{Hz}$ in the inflow direction.  
(b) Plot of $E_{Hz}$ vs.~$t$ at the $y$ location marked in panel (a).}
\end{figure}

A careful determination of when the Hall effect begins to become
important is obtained using a time history plot of the out-of-plane
Hall electric field $E_{Hz} = J_{y} B_{x} / n$ in the inflow direction
through the main X-line, plotted in Fig.~\ref{fig-data2}(a).  (Note,
this cut is in the inflow direction, while Fig.~\ref{fig-data1} is in
the outflow direction.)  The color bar is again stretched.  The plot
clearly shows that $E_{Hz}$ does not contribute during the secondary
island phase.  A cut of $E_{Hz}$ in time, taken at the solid (gray)
line in Fig.~\ref{fig-data2}(a), is plotted in
Fig.~\ref{fig-data2}(b).  The onset time, defined as when $E_{Hz}$
reaches 1\% of its maximum value, is at $t \sim 1780$, the time that
$E$ begins to increase as seen in Fig.~\ref{fig-data}(a).

To emphasize differences between Sweet-Parker with secondary islands
and Hall reconnection, we restart the simulation at $t = 1120$ with
the Hall effect and electron inertia disabled.  The reconnection rate
is plotted as the dot-dashed (red) line in Fig.~\ref{fig-data}(a).
The value reaches $E \sim 0.009$ as the asymptotic upstream field
reaches the dissipation region, in excellent agreement with the
predicted value $E_{SI} \sim E_{SP} \sqrt{N} \sim 0.009$ with $N = 2$
for a single island.  This rate is consistent with the largest scaling
studies done to date \cite{Bhattacharjee09}.  Note, $E$ remains nearly
an order of magnitude slower than Hall reconnection.  Although the
present evidence is based on simulations only up to $S \sim 10^{5}$,
it is clear that secondary island reconnection does not produce the
fastest reconnection rates.

The present results may be relevant for observations of two-phase
reconnection events in the corona.  In observations of flux emergence
\cite{Longcope05}, a slow phase of reconnection preceded an abrupt
transition to a fast phase $\sim 30$ times faster (compare the slopes
in their Fig.~18).  In observations of the contraction of magnetic
loops in an impulsive flare \cite{Liu10}, the contraction velocity
abruptly increased by a factor of $\sim 16$.  It is enticing to
attribute these observations to a transition from resistive secondary
island reconnection at a normalized reconnection rate of $E \sim 0.01$
(consistent with implications of
Refs.~\cite{Daughton09,Bhattacharjee09}) to Hall reconnection $\sim
10$ times faster which occurs abruptly when gyroscales are reached.
The existing level of accuracy of both theory and observations make
such an identification premature, but it remains an exciting
possibility.

Assumptions in this work that require further study include using a
Spitzer resistivity instead of the uniform resistivity employed here,
including Ohmic heating and viscosity, and including Dreicer field
effects, which may be important for the transition to Hall
reconnection.  The simulations have no guide field, but one would be
expected in the corona.  Also, the simulations are two-dimensional;
three-dimensional effects are not included.


The authors gratefully acknowledge support by NSF grant PHY-0902479,
NASA's EPSCoR Research Infrastructure Development Program and a WVU
Faculty Senate Research Grant.  This research used resources at
National Energy Research Scientific Computing Center.  We thank
J.~F.~Drake for helpful conversations.


\end{document}